\documentclass{article}

\usepackage{microtype}
\usepackage{amsmath}
\usepackage{graphicx}
\usepackage{color}
\usepackage{xspace}

\makeatletter
\DeclareRobustCommand\onedot{\futurelet\@let@token\@onedot}
\def\@onedot{\ifx\@let@token.\else.\null\fi\xspace}

\makeatother

\title{Large-Scale Electron Microscopy Image Segmentation in Spark}

\author{Stephen M. Plaza and Stuart E. Berg}

\date{\today}

\begin{document}
\maketitle

\begin{abstract}
The emerging field of connectomics aims to unlock the mysteries of the brain by understanding the connectivity between neurons. To map this connectivity, we acquire thousands of electron microscopy (EM) images with nanometer-scale resolution. After aligning these images, the resulting dataset has the potential to reveal the shapes of neurons and the synaptic connections between them. However, imaging the brain of even a tiny organism like the fruit fly yields terabytes of data. It can take years of manual effort to examine such image volumes and trace their neuronal connections.  One solution is to apply image segmentation algorithms to help automate the tracing tasks.  In this paper, we propose a novel strategy to apply such segmentation on very large datasets that exceed the capacity of a single machine. Our solution is robust to potential segmentation errors which could otherwise severely compromise the quality of the overall segmentation, for example those due to poor classifier generalizability or anomalies in the image dataset.  We implement our algorithms in a Spark application which minimizes disk I/O, and apply them to a few large EM datasets, revealing both their effectiveness and scalability.  We hope this work will encourage external contributions to EM segmentation by providing 1) a flexible plugin architecture that deploys easily on different cluster environments and 2) an in-memory representation of segmentation that could be conducive to new advances.
\end{abstract}

\section{Introduction}
\begin{figure}
\centering
\includegraphics[width=0.8\textwidth]{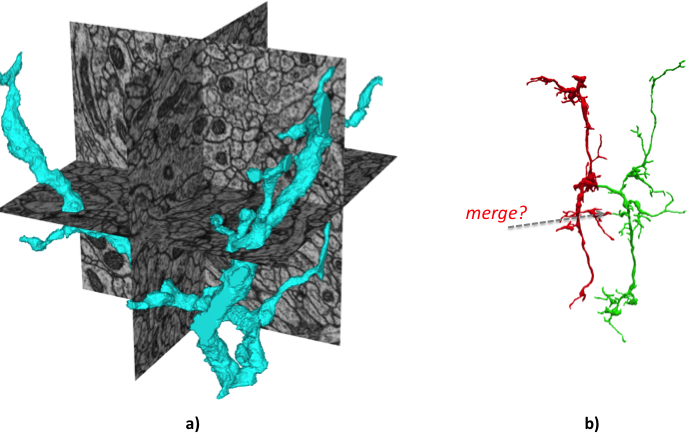}
\caption{\label{fig:neuronex} {\bf Partial neurons
extracted from EM data.} a) Neurons can have intricate
branching and span thousands of images.  b) Single false
merge or false split errors can greatly impact the neuron
shape and its corresponding connectivity.}
\vspace{-1mm}
\end{figure}

The process of extracting or reconstructing neuronal shapes from an EM dataset is laborious \cite{Plaza14}.  As a result, the largest connectomes produced involve only hundreds to thousands of neurons \cite{Sebastian14,Nature13,Takemura15}. Nanometer resolution imaging is necessary to resolve important features of neurons, but as a consequence, a neuron that spans just 50 microns might intersect thousands of image planes.  Also, neurons often exhibit complicated branching patterns.  Figure \ref{fig:neuronex}a shows a small part of an EM dataset and an intersecting neuron.

Image segmentation attempts to reconstruct these shapes automatically, by first identifying voxels belonging to neuronal boundaries and then partitioning the data using these boundaries, such that each partition yields a different neuron.  Despite continual advances in machine learning, the state-of-the-art segmentation approaches \cite{Funke12,jni13,Parag14,Andres,Huang14} still require manual ``proofreading'' \cite{Takemura15}.  Proofreading involves merging falsely split segments and splitting falsely merged ones.  Algorithms are generally biased in favor of false split over false merge errors. This is because manual correction of a falsely split segment is trivial, but manually splitting a falsely merged segment is much more labor intensive, requiring definition of the split boundary.

Recent efforts aim to segment and evaluate large-scale image segmentation \cite{kaynig15,roncal15}.  The general approach involves segmenting small subvolumes and concatenating these subvolumes to form a segmentation for the entire image volume.  Such approaches still do not take the global context of the segmentation problem into account.  Thus, even though state-of-the-art strategies may perform well when compared with small, manually reconstructed ground truth volumes, the quality of their results still suffers on large datasets.

\begin{figure}
\centering
\includegraphics[width=0.8\textwidth]{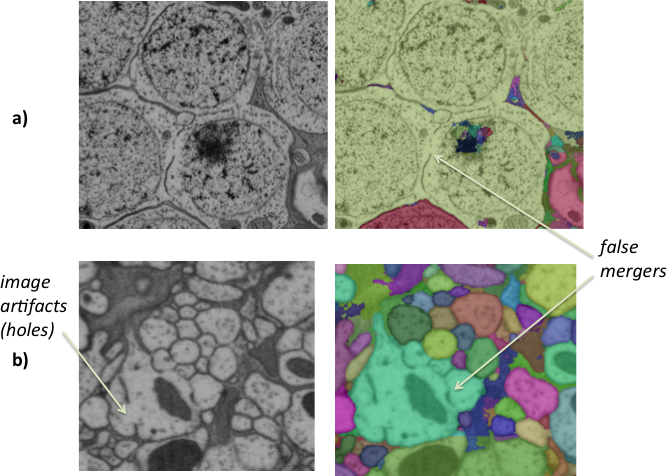}
\caption{\label{fig:badseg} {\bf EM examples
that are hard to segment.} a) This example contains
parts of the neuron not included in the training set.  Poor
classifier generalizability contributes to a bad segmentation
result.  b) Image artifacts such as membrane holes can result in false merging.}
\vspace{-1mm}
\end{figure}

For example, if the segmentation strategy employs a classifier which fails to generalize to all parts of the dataset, false merges may be concentrated in one region of the global dataset (Figure \ref{fig:badseg}), propagating errors throughout the volume. This scenario is especially likely in regions where the original data contains image artifacts.

Even in regions with good classifier performance, sparse errors may not corrupt the local segmentation quality, but the effect of these sparse errors within the larger volume can be catastrophic.  Since neurons typically span vast lengths within a large volume, even relatively sparse errors within a local context can have significant nonlocal consequences. These can severely affect the accuracy of the extracted connectome, as shown in  Figure \ref{fig:neuronex}b.

In addition to these challenges, performing image segmentation on teravoxel-(or higher) scale datasets involves many practical considerations:

\begin{enumerate}
\item Processing a large dataset requires significant compute resources.  Unavoidable crashes (e.g., those due to network outages) or bugs that kill a long running cluster job require costly re-runs.
\item Segmentation algorithms are continually advancing.  It is important to be able to quickly swap and evaluate algorithm components and to partially refine an existing segmentation.
\item Large software frameworks tend to be difficult to understand and deploy by non-experts in diverse compute environments.
\end{enumerate}

In this paper, we introduce an open-source segmentation framework that runs efficiently on large image datasets.  Our framework is implemented in Apache Spark, a fast distributed engine for processing data in parallel \cite{spark}.  Spark is widely adopted on multiple platforms making it easy to deploy in many different compute environments.  We also layer this framework over a volume data-service for accessing and versioning volume data, DVID \cite{dvid}, which allows us to abstract the storage/data layer from the core algorithms.  Other key features of our framework include:

\begin{enumerate}
\item A flexible plugin system to enable easy swapping of segmentation components
\item Check-pointing mechanisms that save partial computations to enable fast rollback and recovery
\item In-memory representations of segmentation to reduce disk I/O
\item A novel segmentation stitching strategy to prevent propagation of local segmentation errors
\end{enumerate}

We evaluated the pipeline on several EM datasets.  We demonstrate the scalability and the efficiency of rollback/recovery on a 400GB volume.  Furthermore, we show the effectiveness of our stitching approach by comparing automatic image segmentation to large ground-truth datasets.

The paper is organized as follows.  In Section \ref{sec:background}, a standard framework for segmentation is described.  We then introduce our large-scale framework in Section \ref{sec:framework} and highlight the key features of the code design in Section \ref{sec:dvidsparkservices}.  Finally, we present results and conclusions.  For reference, the appendix outlines the segmentation plugins available and provides a sample configuration file.

\section{Background}\label{sec:background}
\begin{figure}
\centering
\includegraphics[width=1.0\textwidth]{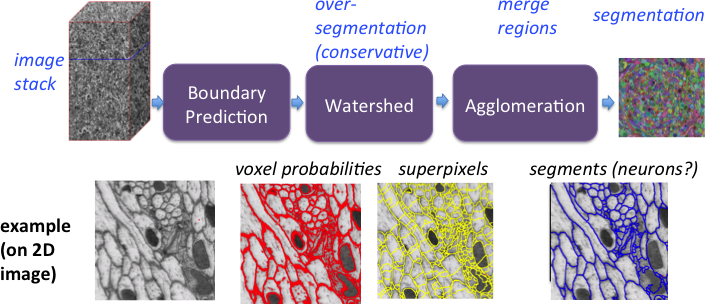}
\caption{\label{fig:basicseg} {\bf Segmentation
workflow for EM dataset.}  The workflow typically involves
voxel prediction, generation of an over-segmentation (using algorithms
like watershed), and agglomerating these segments.}
\vspace{-1mm}
\end{figure}

A general strategy for segmenting an EM volume is shown in Figure \ref{fig:basicseg}.
A classifier is trained to distinguish neuronal membrane boundaries from
the rest of the image.  Applying the classifier to the image volume produces
a boundary probability map.  Typically, some supervoxel generation algorithm such as watershed \cite{watershed} is applied to the data, yielding an oversegmentation of regions within likely boundaries.  Depending on the voxel resolution (nearly isotropic or anisotropic), the previous steps could be applied to either 2D slices or 3D subvolumes.  Because over-segmentation errors are easier to correct than under-segmentation errors, the boundary prediction is often conservative, producing a segmentation with many small fragments.  Clustering or agglomerating this over-segmentation produces the final result \cite{jni13,Parag14}.  In most cases, a neuron is still split into multiple segments.

\begin{figure}
\centering
\includegraphics[width=1.0\textwidth]{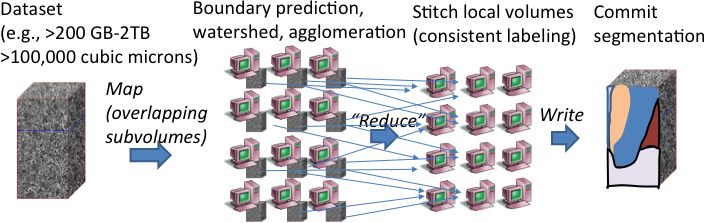}
\caption{\label{fig:largeseg} {\bf Framework
for segmenting large datasets.}  The dataset is split
into several overlapping regions.  Each region is segmented
and the overlapping bodies are stitched together.}
\vspace{-1mm}
\end{figure}

For large datasets, the computation must be partitioned across multiple machines.  Computing the boundary prediction typically requires only local context and is easily parallelized.  While watershed and agglomeration would ideally be computed as global operations across the entire volume, it is reasonable to spatially partition the dataset and apply these algorithms on overlapping blocks.  The shared regions between blocks are then used to merge the blockwise segmentation results together at the end of the pipeline.  This pipeline is shown in Figure \ref{fig:largeseg} and is roughly the strategy used in previous work \cite{kaynig15,roncal15}.  In principle, agglomeration could be done with more global scope by successively partitioning the global
graph that describes the inter-connections between connected components produced by watershed.

\section{Large-scale Segmentation Framework}\label{sec:framework}
We introduce our framework at the algorithm and system level in this section, modeled off of the design in Figure \ref{fig:largeseg}.  In Section \ref{sec:dvidsparkservices}, we will discuss lower-level details.

\subsection{Outlier Resistant Segmentation Stitching}
\begin{figure}
\centering
\includegraphics[width=1.0\textwidth]{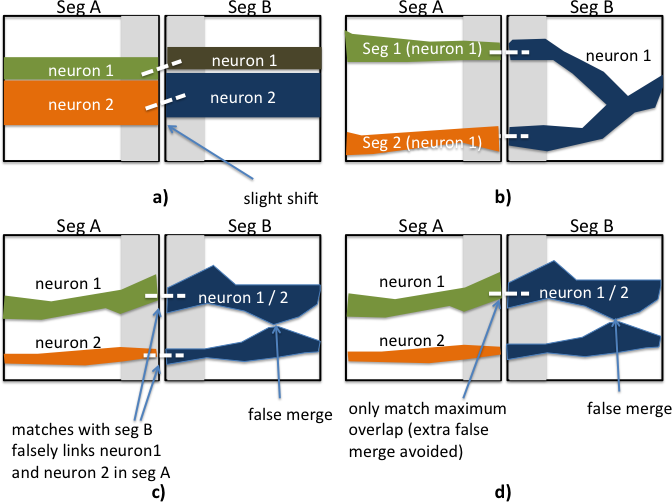}
\caption{\label{fig:stitch} {\bf Scenarios for stitching
neurons based on overlap between subvolumes.}  Four scenarios
are given showing two subvolumes where the overlap is shown
in gray.  a) Slightly different context and algorithm
instability can cause a small shift in the segmentation leading to a non-exact match.  b)  Neuronal branching can link two segments in another subvolume together.  c)  False merging in subvolume $B$ causes false merging in subvolume $A$ due to segment overlap.  d) Conservative matching can avoid propagating false mergers.}
\vspace{-1mm}
\end{figure}

After first generating segmentation on small, overlapping subvolumes, their segments are stitched based on voxel overlap along a subvolume face.  We will first assume that the subvolume segmentation is strictly over-segmented (no false merges).  Even with this simplifying assumption, there will not be 100\% overlap between segments from the same neuron.  In one scenario, slight algorithm instability and differing contexts can cause small boundary shifts in the overlap region between subvolumes of a few pixels.  For example,
Figure \ref{fig:stitch}a shows two neurons running in parallel whose exact boundary varies between subvolumes.  Even with slight overlap $A_2$ should only merge with $B_2$ and not $B_1$.  In another scenario, a neuron can branch near the subvolume border.  Figure \ref{fig:stitch}b shows that segment $A_1$ and $A_2$ should be merged together and joined with segment $B_1$.  Stitching based on any overlap will cause the neuronal segments to be merged correctly, unlike in the first scenario.

A simple stitching rule properly handles both scenarios.  A segment $A_i$ in subvolume $A$ matches a segment $B_j$ in subvolume $B$ (meaning they are in the match set $M_{A_i, B_j}$), if and only if, $A_i$ overlaps with $B_j$ the most or vice versa:

\begin{equation} \label{eq:match}
M_{A_i, B_j} \iff argmax_x (A_i \cap B_x) = j \vee \\
		argmax_y (A_y \cap B_j) = i
\end{equation}

\noindent Applying this rule to all $A_i$ and $B_j$ means that each
segment in $A$ (and $B$) will be linked
to at least one other segment in $B$ (and $A$).  Similarly, we
could define the matching condition
as needing a minimum overlap threshold:

\begin{equation} \label{eq:match2}
M_{A_i, B_j} \iff \frac{|A_i \cap B_j|}{|A_i|} > K \vee \\
		\frac{|A_i \cap B_j|}{|B_j|} > K
\end{equation}

\noindent This would not guarantee one match for each segment.

For an over-segmented volume, these definitions handle most scenarios.
There are some remaining corner cases that could be handled with a few additional
considerations.  For instance, segments that branch outside of the overlap
region can be temporarily split into connected components and each component can
separately seek a match.  Segments with a large overlap in absolute voxel count could be matched even if the above conditions fail to hold.

Complications arise in practice because of false merging.  When two neuronal segments in a substack are falsely joined, these errors can propagate to the substack boundary.  Given the complexity of using higher-level biological context to guide segmentation and stitching, it not easy to tell whether branching that occurs at a substack face is the result of error or biology.  Figure \ref{fig:stitch}c shows an example where two neurons are falsely merged due to an image artifact, causing the simple stitching rule to result in more errors.  A single substack with bad image quality or poor segmentation for any other reason could potentially propagate errors throughout the entire volume.  The effects are more dramatic for larger volumes since a single neuron spans multiple subvolumes, and there are more opportunities for false merging.

Future work will exploit shape priors to eliminate error propagation and hopefully identify the sources of the false merges.  For now, we introduce a straightforward strategy that admits many matches but conservatively avoids ones that are the most dangerous.  Namely, we avoid multiplying errors by
eliminating matches that bridge multiple segments along a single subvolume face.

For a given subvolume face, we specify that a match between two segments implies that a segment in $A$ and a segment in $B$ each only have one match:

\begin{equation} \label{eq:conservative}
M_{A_i,B_j} \implies \exists_{=1} A_x \in A: M_{A_x, B_j}\ \wedge \ \exists_{=1} B_y \in B: M_{A_i, B_y}
\end{equation}

\noindent We implement a couple of heuristics to satisfy the above constraint if $A_i$ or $B_j$ has multiple matches according to
Equation \ref{eq:match}.  In one approach, we can satisfy the condition by changing $\vee$ to $\wedge$ in Equation \ref{eq:match}.  For example, if a neuron branches, we potentially choose just one branch, where there is largest mutual overlap.  This is shown in Figure \ref{fig:stitch}d.  We also consider an even more conservative approach, by changing the $\vee$ to $\wedge$
in Equation \ref{eq:match2} with $K > 0.5$.

\subsection{Iterative Segmentation}
Image segmentation over a large dataset can be very time consuming, requiring
significant cluster resources.  This computational load increases the likelihood
that the segmentation job will fail, or, even when successful, will be invalidated by newer, better
segmentation results.  First, a long-running job is more vulnerable to network outages or
other events that might disrupt the computation on a shared compute resource.  Second, any software
bugs might be uncovered only after significant portion of the computation has already completed successfully.  In both cases, a potentially costly rerun of the pipeline may be necessary, but unnecessary recomputation of satisfactory results should be avoided.

\begin{figure}
\centering
\includegraphics[width=1.0\textwidth]{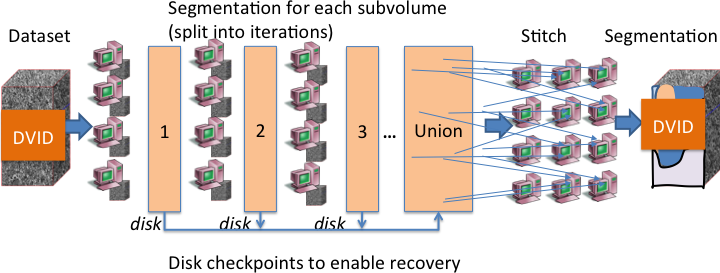}
\caption{\label{fig:robust} {\bf Our robust segmentation
framework.}  Disjoint subsets of subvolumes are processed
and checkpointed to allow recovery from any software
crash.  Note that the dataset and final segmentation is
stored using DVID.}
\vspace{-1mm}
\end{figure}

Figure \ref{fig:robust} illustrates our strategy to make the segmentation pipeline more robust to failures.  Specifically, we focus on the subvolume segmentation component since each task is disjoint and the other parts of the pipeline are comparatively less time consuming.  Our solution
is to divide the set of subvolumes spanning the large dataset across multiple iterations.  In each iteration, a disjoint subset of the subvolumes is processed. The segmentation labels for each subvolume in this procedure are compressed using lossless compression (such as lz4), and serialized to disk.  If there are any unexpected errors in the middle of the job, the pipeline will automatically rollback to the previously computed result ({\em i.e.,} the most recently completed iteration).
We note in the results that the high compressability of these label volumes results in a minimal memory and I/O footprint for these checkpoints.

As mentioned earlier, there is another risk to time consuming segmentation runs.  Namely, it is often desirable to run multiple algorithms and to make successive refinements to the segmentation approach.  However, in our application domain, the segmentation is continually refined and examined and it is undesirable to wait several months to begin this process.  To address these issues, we have the
option to retain previously examined (proofread) results across future segmentations.  Specifically,
we provide the option to preserve a set of segment IDs.  These segments are masked out of the image,
so that the watershed algorithm floods around these regions.

\subsection{Image Data Access}
Our segmentation approach requires efficient access to subsets of a large dataset, which poses many infrastructural challenges.  As EM datasets extend to the 100TBs and beyond, storing this data in a single file on a single machine, such as in HDF5 format, is limiting in bandwidth.  Furthermore, distributing data using a customized system of sharding file blobs might
not generalize across different compute environments and also fails to reuse ongoing efforts outside
of connectomics in large-scale database technology.  These issues are complicated by the desire to
also keep track of segmentation changes due to proofreading or revisions in the segmentation algorithms.

To handle these issues, we adopt the Distributed, Versioned, Image-Oriented Dataservice (DVID) \cite{dvid} to access our large image volume data.  DVID allows one to access large volumetric data through a RESTful API and track changes in label data using a \textit{git}-like model of commits and branches.  By satisfying the service interface, we can isolate the details of data storage from our segmentation services.  In particular, we fetch subvolumes through DVID and write segmentation label data back through DVID.

\section{Open-source Spark Implementation: \\ DVIDSparkServices} \label{sec:dvidsparkservices}
We implement the segmentation framework {\em DVIDSparkServices} using Apache Spark, which allows us to manipulate these large
label volumes in shared memory across several cluster nodes.  While the subvolume segmentation
part of the pipeline requires minimal inter-process communication and might not benefit much from
shared memory dataset representation, we believed Spark to be an ideal framework for several reasons:

\begin{itemize}
\item Spark supports several primitives for working on large distributed datasets which support more structured semantics than more traditional approaches, such as ad-hoc batch scripts, would provide.

\item Spark is supported on many different cluster compute environments.

\item An in-memory representation for large label data will empower future algorithms to use high-level, global context to improve segmentation quality.

\end{itemize}
 
The next subsection details the framework and its use of Spark primitives.  Then we highlight
a plugin system to enable outside contributors to flexibly swap algorithms for different parts of the pipeline.
 
\subsection{Architecture} 
DVIDSparkServices contains several workflows for working with the DVID dataservice. Each workflow is designed as a custom Spark Application, written in Python using PySpark.  Access to and from
DVID is controlled through the sparkdvid module.

Our segmentation pipeline framework is one such workflow. The key stages of the pipeline and technical details are listed below, with a description of the Spark primitive used in each stage.

\begin{enumerate}
\item Logically split the dataset space by defining overlapping bounding boxes (one per subvolume) which, collectively, provide coverage over a region of interest (ROI).  (primitive: parallelize)

\item Map each subvolume's bounding box into a grayscale subvolume. This requires reading subvolumes in parallel using DVID.  (primitive: map)

\item Divide the subvolumes into groups, each of which will be processed in a separate iteration.
\begin{enumerate}

\item Map each grayscale subvolume into a volume of probabilities using some voxel-level classifier.  (primitive: map)

\item Map each probability volume into an over-segmentation of supervoxels.  (primitive: map)

\item Map each over-segmentation into a final subvolume segmentation via agglomeration.  (primitive: map)

\item Optionally serialize the subvolume segmentation to disk for checkpointing. (primitive: saveAsObject)

\end{enumerate}

\item Extract overlapping regions from each pair of neighboring subvolumes.  (primitive: flatMap)

\item Group subvolume overlap regions together.  This requires data to be shuffled throughout the network.  (primitive: groupByKey\footnote{Note that groupByKey is okay since each subvolume region is by construction in different partitions anyway.})

\item Map each overlap region into a set of matches.  Return these matches to the driver, transitively resolve the matches across the subvolumes, and broadcast to each subvolume.  (primitives: map, collect, broadcast)

\item Apply the matches to each subvolume.  (primitive: map)

\item Write segmentation to DVID through sparkdvid.  (primitive: foreach)
\end{enumerate}

\begin{figure}
\centering
\includegraphics[width=1.0\textwidth]{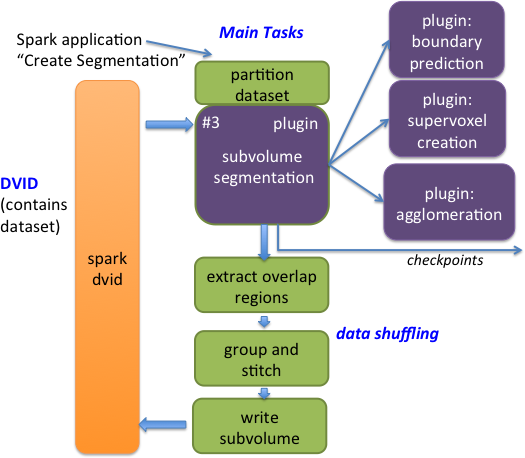}
\caption{\label{fig:sparkarch} {\bf Main components of
our Spark framework.}  The segmentation framework
is a Spark application that transforms the raw image
dataset into a segmentation.  {\tt sparkdvid} handles communication
with DVID.  The subvolume segmentation can be defined
by a custom plugin.  The default plugin allows customization of boundary prediction, supervoxel creation, and agglomeration.  Subvolume stitching is the only major
operation that causes data to be shuffled across the network.}
\vspace{-1mm}
\end{figure}

The main points are emphasized in Figure \ref{fig:sparkarch}.  Most operations require little
communication from the driver.  The primary exception is extracting the overlapping regions of a subvolume and matching them, which requires data to be moved throughout the Spark cluster.  Given the relatively small size of the overlap region compared to the subvolume (close to 10\% of the volume for each subvolume face in our experiments) and the high compressibility of the labels, there is not much data that needs to be moved around.  Step
\#3 executes the core segmentation algorithms.

\subsection{Plugin System}
Our framework is intended to flexibly adopt new algorithms and tailor
solutions to specific application domains.  To do this, we provide several plugin 
interfaces to our segmentation framework, which can be specified when calling the workflow
using a configuration file (see Appendix \ref{sec:config} for an example configuration file).  The plugin
is defined in python and can call any code necessary to satisfy the given interface.

Figure \ref{fig:sparkarch} highlights the main
plugin options available currently in the segmentation framework.
At the top-level, the entire segmentation module which takes grayscale volume data and returns labels
for a given subvolume can be overwritten.  The default segmentation plugin itself allows fine-grain plugin-based control over each stage of the pipeline. The voxel prediction, supervoxel creation, and agglomeration can all be customized and implemented with an alternative python function.  We briefly describe available plugins in Appendix \ref{sec:plugin}.

\section{Results}
The package DVIDSparkServices implements our segmentation workflow and is available on github (https://github.com/janelia-flyem/DVIDSparkServices).  We evaluate this code both for its robustness
at scale and the effectiveness of its stitching strategies.  In all examples, we read and write
data through DVID running over an embedded levelDB database \cite{leveldb} on a single machine.  Therefore,
read and write throughput will be a be limited by the I/O capacity of the database server.  We are using the Ilastik \cite{ilastik} plugin for boundary
prediction and the NeuroProof \cite{Parag14} plugin for agglomeration.  The subvolume size used is 512x512x512, plus an additional 20 pixel border on all faces to create overlap with adjacent subvolumes.

\subsection{Large-scale Segmentation of Fly Optic Lobe}
\begin{figure}
\centering
\includegraphics[width=1.0\textwidth]{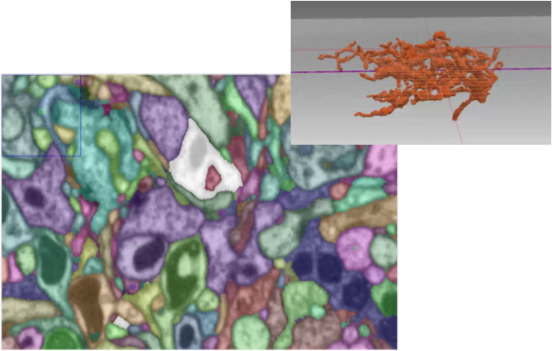}
\caption{\label{fig:fib19seg} {\bf One image plane from the segmentation
of a large optic lobe dataset.} The false coloring depicts the segmented
neurons and the 3D neuron is a previously proofread body that was untouched
during this re-segmentation.}
\vspace{-1mm}
\end{figure}

We applied image segmentation on a portion of the fly optic lobe around
232,000 cubic microns in size imaged at approximately 8x8x8nm resolution using
FIB-SEM \cite{fib}.  This encompasses 453 GB of data or 3,375
subvolumes based on our partitioning.  The segmentation preserves
pre-existing proofread labels.  That is, only as-yet unproofread voxels are replaced with automated segmentation results.  Figure \ref{fig:fib19seg} shows an example
segmented slice.

\begin{table}
\centering
\begin{tabular}{|l|r|}\hline
Task & Time (hours) \\ \hline \hline
subvolume ingestion and segmentation & 42  \\
stitching, remapping and shuffling & 1  \\
subvolume writes & 20  \\ \hline
total & 63 \\
\hline \end{tabular}
\caption{
\label{tab:runtime}
{\bf Rough breakdown of time to run segmentation on a large
portion of the optic lobe dataset}.  The subvolume segmentation
involves seven checkpoint iterations.  The main component of runtime
is the subvolume voxel prediction.  We expect significant improvements
to subvolume writes if we use a distributed back-end database behind
DVID.}
\vspace{-1mm}
\end{table}

Table \ref{tab:runtime} shows the runtime
for segmenting this dataset on a cluster of 32 machines with 16 cores
and 90 GB memory each.  The majority of the runtime is spent in subvolume segmentation.  Only a small percentage of time is required to stitch the
subvolumes together.  As previously noted, we ran our experiment
against a single server datastore and therefore the read/write bandwidth
is limited.  With a distributed datastore behind DVID, it would be possible
to ingest compressed segmented data much faster.

We ran the segmentation using the checkpoint system by dividing the 3,375 subvolumes into seven iterations.  If we were to store all the subvolumes
using uncompressed 64bit labels, we would require 4.5 TB.  Using lz4 compression, the checkpoint system required only 95 GB.  Restoring the checkpoint after all subvolume segmentation was complete took
only around 30 seconds.

\subsection{Evaluation of Subvolume Stitching Algorithms}
To evaluate the stitching algorithm, we compare stitching strategies
on two different datasets.  We evaluate a conservative matching algorithm using Equations \ref{eq:match} and \ref{eq:conservative} and a more aggressive strategy that additionally considers matches which satisfy Equation \ref{eq:match2}.\footnote{The actual implementation uses additional constraints to avoid matching segments with negligible overlap.}  Since we are evaluating
segmentation quality, we require ground truth data.  The difficulty of generating ground truth limits the number of possible comparisons.  Fortunately, a large ground truth dataset of 27,000 cubic microns is available in the optic lobe medulla \cite{Takemura15}.  We also evaluate segmentation on an unpublished dataset from the fly mushroom body (MB) that is 5,250 cubic microns.

\begin{table}
\centering
\begin{tabular}{|l|r|r|r||r|r|r|}\hline
stitching & \multicolumn{3}{|c||}{medulla} & \multicolumn{3}{|c|}{mushroom body} \\ 
strategy & f.merge VI & f.split VI & VI & f.merge VI & f.split VI & VI \\ \hline \hline
no stitch & 0.05  & 4.36 & 4.41  & 0.05 & 4.36 & 4.41 \\
conservative & 0.68 & 1.60 & {\bf 2.28} & 0.15 & 2.27 & 2.42\\
aggressive & 2.76 & 1.10 & 3.86 & 0.24 & 2.08 & {\bf 2.31} \\
\hline \end{tabular}
\caption{
\label{tab:vi}
{\bf Evaluation of segmentation using three stitching strategies}.  Our
stitching strategies result in lower total VI than no stitching.  However,
more aggressive stitching can result in several false mergers (high false merge VI) as seen in the medulla sample.}
\vspace{-1mm}
\end{table}

\begin{figure}
\centering
\includegraphics[width=1.0\textwidth]{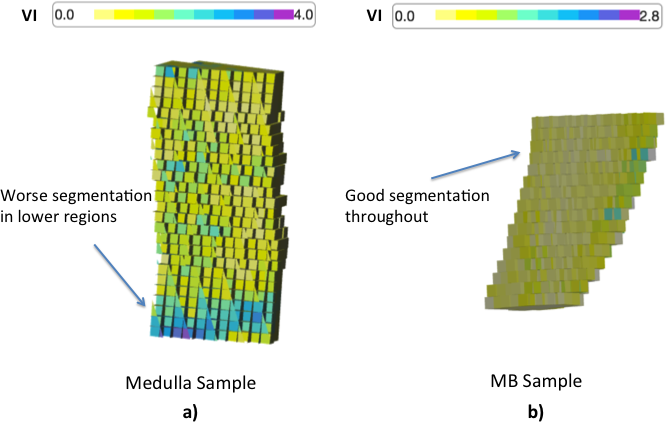}
\caption{\label{fig:heatmap} {\bf Segmentation quality of aggressive stitching throughout a large dataset.}  The VI metric is computed on small regions to provide a heatmap indicating how segmentation quality
varies as a function of location.  Connected components is run in each subvolume to compare against the ground truth.  (It is possible that bad false merging outside of a given subvolume is not fixable by connected components within the region.) a) The larger medulla sample has poor segmentation in the lower regions which results in a lot of false merging.  b) The smaller MB sample has more uniform segmentation quality.}
\vspace{-1mm}
\end{figure}

Table \ref{tab:vi} shows the quality of segmentation using the conservative and straightforward matching strategy over the two datasets.  The similarity is scored using Variation of Information (VI) \cite{vi}.  This metric allows us to 
decompose the similarity into a false merge ({\tt f.merge}) and false split ({\tt f.split}) score.  A higher number indicates more errors.  In both datasets, we see that both stitching strategies results in better total VI than
no stitching.  Conservative stitching results in more false splits than aggressive stitching.  However, for the medulla dataset,
the conservative stitching produces significantly less false merge errors and therefore a better overall segmentation.  In general, false merges are harder
to fix manually.

A parameter controls the aggressiveness of the stitching procedure.  The setting should be chosen based on the expected quality of the subvolume segmentation results.  In general, a large dataset probably
increases the odds that there is some anomaly or problem that could cause false
merging.  (The medulla dataset is considerably larger than the MB dataset.)
Figure \ref{fig:heatmap} shows the false merge VI in small subregions of the large volume as a heat map.  Notice that there are few areas toward the bottom of the medulla dataset that contains more false merge errors.  Despite the segmentation being
good in many parts of the dataset, the overall similarity is compromised.

\section{Conclusions}
We introduce a large-scale, open-source EM segmentation infrastructure implemented in Apache Spark.  We show that our segmentation can robustly
segment large datasets by avoiding the propagation of bad false mergers
and efficiently maintaining checkpoints.  Our implementation allows
custom plugins for different parts of the segmentation pipeline.  We have
tested our system on both our internal compute cluster and on
Google Compute Engine.  The use of Spark will allow the solution
to easily port to different compute environments.

Future work entails exploiting the more global context afforded by our system to guide
new segmentation and stitching algorithms.  The ability to store
a large segmentation in memory across several machines will enable
novel strategies.  We also plan to leverage ongoing research to
add different storage backends to DVID.  By using a distributed database
backend, read and write bottlenecks will be greatly alleviated.

{\small
\vspace{2mm}
\noindent \textbf{Acknowledgements:} We would like to thank the FlyEM project
team at Janelia Research campus.  Zhiyuan Lu, Harald Hess, and Shan Xu prepared fly samples and created the image datasets.  Pat Rivlin, Shin-ya Takemura, and
the FlyEM proofreading team provided biological guidance and the groundtruthing effort. Toufiq Parag provided image segmentation insights and helped in tuning segmentation performance.  Bill Katz implemented DVID API that used in our segmentation system.}

\newpage
\appendix

\begin{center}
\vspace{2mm}
{\bf APPENDIX}
\end{center}

\section{Overview of Segmentation Workflow Classes} \label{sec:plugin}
The segmentation workflow in the DVIDSparkServices package is defined by the {\tt CreateSegmentation} class.  This class implements the pre-segmentation logic for the following steps:

\begin{enumerate}
\item dividing the region of interest into logical subvolumes
\item grouping subvolumes into iterations
\item mapping subvolumes to grayscale data
\end{enumerate}

\noindent and the post-segmentation logic for the following steps:

\begin{enumerate}
\item serializing the subvolume segmentation results
\item extracting the segmentation pixels for overlapping regions between substacks and finding matches between adjacent subvolume segments
\item uploading the final stitched segmentation results to DVID
\end{enumerate}

The work of actually segmenting each subvolume block and stitching the blocks together is performed in a separate class, {\tt Segmentor}.  The Segmentor class implements two methods: {\tt segment()} and {\tt stitch()}.

\subsection{Customizing the behavior of Segmentor.segment()}

The {\tt segment()} function is implemented in three steps, each of which can be customized by implementing a python function, which will be called at the appropriate time:

\begin{itemize}
\item predict-voxels
\item create-supervoxels
\item agglomerate-supervoxels
\end{itemize}

To customize one of these steps, a developer must simply implement a python function with the appropriate signature, and provide the name of that function in the configuration JSON file as described below.  The logic for obtaining and uncompressing the input data to each function is handled in {\tt Segmentor}, so the custom python functions at each stage merely deal with plain numpy arrays.

For example, an extremely simple method for producing voxel-wise membrane ``predictions'' might be the following:

\begin{verbatim}
# mymodule.py
def predict_via_threshold(grayscale_vol, mask_vol, threshold):
    # This function assumes dark pixels are definitely membranes.
    # All dark pixels are returned as 1.0, and everything else 0.0.
    return (grayscale_vol < threshold).astype(numpy.float32)
\end{verbatim}

The {\tt Segmentor} class can be instructed to call this function during the ``predict-voxels'' step via the config file.  Parameters such as ``threshold'' should also be specified in the config file, as shown in the example below.

\begin{verbatim}
{
  "options": {
    "segmentor": {
      "class" : "DVIDSparkServices.reconutils.Segmentor",
      "configuration": {
        "predict-voxels" : {
          "function": "mymodule.predict_via_threshold",
          "parameters": {
            "threshold": 30
          }
        }
      }
    },
    ... additional configuration settings omitted ...
  }
}

\end{verbatim}

As demonstrated above, each custom function must accept a specific set of required arguments, and any number of optional keyword arguments.  The following describes each customizable function's purpose, along with its required arguments.

\begin{itemize}

\item {\bf predict-voxels}
 Given a grayscale volume and a boolean mask of 'background' pixels for which voxel predictions are not needed, return a volume of predictions (range: 0.0 - 1.0, dtype: float32) indicating the probability of the presence of a membrane at each pixel.  Additional channels representing other probability classes may be optionally appended to the first channel.
  Required arguments: {\tt grayscale\_vol}, {\tt mask\_vol}

\item {\bf create-supervoxels}
  Given the prediction volume from the ``predict-voxels'' step and a background mask volume, produce an oversegmentation of the volume into supervoxels.  The supervoxels must not bleed into any background regions, as indicated by the background mask.
  Required arguments: {\tt prediction\_vol}, {\tt mask\_vol}

\item {\bf agglomerate-supervoxels}
  Given the prediction volume from the ``predict-voxels'' step and the oversegmentation from the ``create-supervoxels'' step, produce a segmentation volume.
  Required arguments: {\tt predictions}, {\tt supervoxels}
\end{itemize}

\subsection{Built-in Segmentation Functions}

As described above, each stage of the {\tt Segmentor.segment()} method can be customized, but there is no need to implement your own functions for every stage from scratch.  The DVIDSparkServices package already includes built-in functions which can be used for each stage.  Each of these is defined within the {\tt DVIDSparkServices.reconutils.plugins} namespace.

\begin{itemize}
\item predict-voxels step
\begin{itemize}
\item {\tt ilastik\_predict\_with\_array()}
    
Performs voxel prediction using a trained ilastik Pixel Classification project file (.ilp).

\item {\tt two\_stage\_voxel\_predictions()}

Run a two-stage voxel prediction using two trained ilastik Pixel Classification project files.  The output of the first stage will be saved to a temporary location on disk and used as input to the second stage.

\item {\tt naive\_membrane\_predictions()}

Implements an extremely naive heuristic for membrane probabilities by simply inverting the grayscale data.  This function is mostly intended for testing purposes, but for extremely clean data, this might be sufficient.
\end{itemize}

\item create-supervoxels step
\begin{itemize}
\item {\tt seeded\_watershed()}

Computes a seeded watershed over the membrane prediction volume. The seeds are generated by simply thresholding the prediction volume.
\end{itemize}

\item agglomerate-supervoxels step
\begin{itemize}
\item {\tt neuroproof\_agglomerate()}

Agglomerates the oversegmentation image using a trained NeuroProof classifier file.
\end{itemize}

\end{itemize}

\newpage

\section{Example JSON Configuration File}
\label{sec:config}
\begin{verbatim}
{ "dvid-info": {
    "dvid-server": "127.0.0.1:8000",
    "uuid": "abcde12345",
    "segmentation-name": "my-segmentation-result",
    "roi": "my-predefined-region-of-interest",
    "grayscale": "grayscale"
  },
  "options": {
    "segmentor": {
      "class" : "DVIDSparkServices.reconutils.Segmentor",
      "configuration": {
        "predict-voxels" : {
          "function":
            "DVIDSparkServices.reconutils.plugins.ilastik_predict_with_array",
          "parameters": {
            "ilp_path": "/path/to/my/trained-membrane-predictor.ilp",
            "LAZYFLOW_THREADS": 1,
            "LAZYFLOW_TOTAL_RAM_MB": 1024
          }
        },
        "create-supervoxels" : {
          "function": "DVIDSparkServices.reconutils.plugins.seeded_watershed",
          "parameters": {
            "boundary_channel": 0,
            "seed_threshold": 0.01,
            "minSegmentSize": 300,
            "seed_size": 5
          }
        },
        "agglomerate-supervoxels" : {
          "function":
            "DVIDSparkServices.reconutils.plugins.neuroproof_agglomerate",
          "parameters": {
            "classifier": { "path" : "/path/to/my/np-agglom.xml"},
            "threshold": 0.2,
            "mitochannel": 2
          }
        }
      }
    },
    "stitch-algorithm" : "medium",
    "chunk-size": 512
  }
}
\end{verbatim}
\vspace{-2mm}

\begin{thebibliography}{1}
\bibitem{Andres}
B. Andres, {\em et al.}, ``3D segmentation of SBFSEM images of neuropil by a graphical model over supervoxel boundaries,'' {\em Med. Image Anal}, 2012, pp. 796–805.

\bibitem{watershed}
S. Beucher, F. Meyer, ``The morphological approach to segmentation : the watershed transformation,'' {\em Mathematical
Morphology in Image Processing} 1993, pp. 433–481.

\bibitem{Funke12}
J. Funke, B. Andres, F. Hamprecht, A. Cardona, M. Cook, ``Efficient automatic 3D-reconstruction of branching neurons from EM data.'' {\em Proc. IEEE Conference on Computer Vision and Pattern Recognition}, 2012, pp. 1004–1011.

\bibitem{Huang14}
G. Huang, V. Jain, ``Deep and wide multiscale recursive networks for robust image labeling,'' {\em CoRR}, 2013.

\bibitem{dvid}
W. Katz, ``Distributed, Versioned, Image-oriented Dataservice (DVID),'' {\tt http://github.com/janelia-flyem/dvid}

\bibitem{kaynig15}
V. Kaynig, {\em et al.}, ``Large-scale automatic reconstruction of neuronal processes from electron microscopy images,'' {\em Medical Image Analysis}, 2015, 22(1), pp. 77-88.

\bibitem{Sebastian14}
J. Kim,	 M. Greene,	 A. Zlateski, K. Lee, M. Richardson, ``Space–time wiring specificity supports direction selectivity in the retina,'' {\em Nature}, May 2014, pp. 331-336. 

\bibitem{fib}
G. Knott, H. Marchman, D. Wall, B. Lich, ``Serial section scanning electron microscopy of adult brain tissue using focused ion beam milling,'' {\em J. Neurosci}, 2008, pp. 2959-2964.

\bibitem{vi}
M. Meila, ``Comparing Clusterings. '' {\em Proceedings of the Sixteenth
Annual Conference on Computational Learning Theory}, 2003, Springer.

\bibitem{jni13}
J. Nunez-Iglesias, R. Kennedy, T. Parag, J. Shi, D. Chklovskii, ``Machine Learning of Hierarchical Clustering to Segment 2D and 3D Images,'' {\em PLoS ONE}, August 2013, 8(8): e71715. doi: 10.1371/journal.pone.0071715.

\bibitem{raveler}
D. Olbris, P. Winston, S. Plaza, M. Bolstad, P. Rivlin, L. Scheffer, D. Chklovskii, ``Raveler: A proofreading tool for EM reconstruction,'' {\em unpublished}, 2016.

\bibitem{Parag14}
T. Parag, A. Chakraborty, S. Plaza, `A Context-aware Delayed Agglomeration Framework for EM Segmentation`Analysis Paper,'' {\em ArXiV}, June 2014.

\bibitem{Plaza14}
S. Plaza, L. Scheffer, D. Chklovskii, ``Toward Large-Scale Connectome Reconstructions," {\em Current Opinion in Neurobiology}, April 2014, pp. 201-210.

\bibitem{Plaza12}
S. Plaza, L. Scheffer, M. Saunders, ``Minimizing Manual Image Segmentation Turn-Around Time for Neuronal Reconstruction by Embracing Uncertainty,'' {\em PLoS ONE}, September 2012, 7(9): e44448. doi: 10.1371/journal.pone.0044448

\bibitem{roncal15}
W. Roncal, {\em et al.}, ``An automated images-to-graphs framework for high resolution connectomics,'' {\em Frontiers in Neuroinformatics}, 2015, 9:20. doi: 10.3389/fninf.2015.00020.

\bibitem{ilastik}
C. Sommer, C. Straehle, U. Koethe, F. Hamprecht, ``ilastik: interactive learning and segmentation toolkit,'' {\em Proc. IEEE International Symposium on Biomedical Imaging}, 2011, pp. 230–233.

\bibitem{Nature13}
S. Takemura, A. Bharioke, Z. Lu, A. Nern, S. Vitaladevuni, {\em et al}, ``A visual motion detection circuit suggested by Drosophila connectomics,'' {\em Nature}, 2013, pp. 175-181.

\bibitem{Takemura15}
S. Takemura, {\em et al.}, ``Synaptic circuits and their variations within different columns in the visual system of Drosophila,'' {\em PNAS}, 2015, 112, 44, pp. 13711-13716.

\bibitem{neutu}
T. Zhao, ``NeuTu,'' {\tt http://github.com/janelia-flyem/NeuTu}

\bibitem{spark}
M. Zaharia, M. Chowdhury, M. Franklin, S. Shenker, I. Stoica, ``Spark: cluster computing with working sets,'' {\em HotCloud}, 2010, pp. 10-10.

\bibitem{leveldb}
''LevelDB: a fast and lightweight key/value database,"" {\tt https://github.com/google/leveldb}.

\end{thebibliography}
\end{document}